\documentstyle[12pt,epsfig]{article}
\begin{document}
\date{}
\bigskip
\hspace*{\fill}
\vbox{\baselineskip12pt \hbox{UCSBTH-97-22}\hbox{hep-th/9711106}}
\bigskip\bigskip\bigskip

\begin{center}
{\Large {\bf Born-Infeld Strings Tunneling to D-branes}}

\vspace{48pt}

Roberto Emparan\footnote{Address after Jan 1, 1998: Dept.\ of 
Mathematical Sciences, University of Durham, Durham DH1 3LE, UK.}

\vspace{12pt}

{\sl Department of Physics}\\
{\sl University of California}\\
{\sl Santa Barbara, CA 93106, USA}\\
{\it emparan@cosmic1.physics.ucsb.edu}

\vspace{72pt}

{\bf Abstract}
\end{center}
\begin{quote}{%\small
Recent work on solutions to the Born-Infeld theory used
to describe D-branes suggests that fundamental strings can be viewed, 
in a certain limit, 
as D-branes whose worldvolumes have collapsed to string-like 
configurations. Here we address the possibility of undoing this
collapse, by inducing the string to tunnel to an extended brane
configuration.
Using semiclassical methods we argue that, by putting one of these 
Born-Infeld strings in the background
of a uniform 4-form RR field strength, the string can nucleate a 
spheroidal bulge of D2-brane, or, if the string is wrapped around
a small enough circle, it can tunnel to a toroidal D2-brane.
This process can also be interpreted in terms of the M2-brane.
We also address the extension to other D$p$-branes, and discuss
the range of validity of the approximations involved.
}\end{quote}

\medskip

\newpage

\section{Introduction}

The usual logical (and historical) path leading to D-branes starts 
from their recognition as the extended objects 
where open strings can stick their endpoints 
\cite{joe}. Perturbative string theory then allows 
to deduce that, at low energies, D-brane dynamics is 
described by a non-linear Born-Infeld (BI) theory on the 
worldvolume of the brane \cite{leigh}.
Recently it has been found that, to some extent, this path can
be traversed in the opposite direction:
the BI theory by itself reveals the existence of 
string-like excitations that end on (or rather sprout from) 
the worldvolume of the brane \cite{cama,bion}. 
These have the same tension and charge
as a fundamental string. Moreover, at weak coupling, they
scatter waves off the brane as expected from an open string with
Dirichlet boundary conditions \cite{cama}. Further evidence
for the identification of the Born-Infeld string with a fundamental
string has been recently given in \cite{lpt}.
It is then conceivable that this description of strings using
Born-Infeld theory may be useful in some limit where
gravitational (and other closed string) effects can be neglected, 
while keeping the effects due to open strings with Dirichlet 
boundaries. Such a field theory description
of the string may be provide a handle in regimes beyond the scope of
perturbative string theory.

Indeed, the picture above suggests to view a string as a $p$-brane 
($p\geq 2$) with a worldvolume in which 
$(p-1)$ spatial directions have collapsed to zero size. Note 
that we do not want the $(p-1)$ collapsing directions 
of the brane to wrap any non-trivial $(p-1)$-cycle. This is because
we do not want the collapsing D-brane to 
carry any net conserved RR charge. For instance, 
we can start from a brane in flat space, with a worldvolume where
the topology of the spatial sections is ${\bf R}\times S^{p-1}$,
and then we let the $S^{p-1}$ shrink down to zero size 
(as indeed it must do when
there is nothing to oppose the brane tension). 
The axion charge of the string is provided by electric flux in the 
worldvolume of the brane running along the non-collapsed (string) 
direction, and in this sense the starting point is a D$p$-brane with 
a fundamental string dissolved in its worldvolume. 
It should be understood that the different D$p$-branes all 
collapse to the same fundamental string. The different representations
correspond to the sector of the string dynamics associated to 
modes of open strings with Dirichlet boundary conditions
on a $(9-p)$-dimensional manifold.

Now, if the string can be viewed as a collapsed brane, 
can it somehow exhibit its D-braneous
aspects? One could envisage the possibility that a string, by 
application of external forces, may expand into a D-brane.
Consider, in particular,
a given background RR $(p+2)$-form field strength. 
It is well known
that a D$p$-brane couples minimally to such field, just like 
electrons couple to electric Maxwell fields. As a matter of fact, 
this coupling leads to a higher
dimensional analog of the Schwinger pair creation process:
spherical $p$-branes can be spontaneously 
nucleated in a uniform field background, 
in a typical instanton-mediated process \cite{teit,dggh}.
After tunneling, the branes expand with uniform acceleration 
under the pull of the field.\footnote{It may be useful to recall
that the effect of a field strength $H_{01\dots p+1}$ (uniform 
over spacetime)
on a spherical $p$-brane extended along the directions 
$X^1,\dots,X^{p+1}$, is a uniform radial, outward or inward, pull.}

The precise question we address in this paper is whether a 
string, pictured as a collapsed brane, can 
respond to an external uniform field strength by similarly tunneling
to a configuration where the brane (or part of it)
is not collapsed. Since such a decay process cannot be seen in string
perturbation theory, this is an instance where the BI description
of the string is useful. 

We will perform the analysis explicitly only
for $p=2$, i.e., the type IIA string as a collapsed D2-brane. 
As we will see below, 
this is equivalent to the conventional view of the string as
an eleven-dimensional membrane wrapped around a circle. 
There are no technical difficulties in the 
generalization to higher D$p$-branes, but the conclusions turn out 
to be qualitatively different. This will be discussed near the end.

A major point of concern with this approach is the use of Born-Infeld
theory to describe brane configurations with large curvatures 
($ \sim \partial^2 X$) or worldvolume field gradients ($\partial 
{\cal F}$). Born-Infeld theory is only the leading approximation to the
open string effective action, and as such, it would be expected to 
receive corrections at scales of order $\alpha'$. Indeed, 
this makes the results in \cite{cama,lpt} look very surprising. 
Even though the ``spike'' solution
has been shown in \cite{thorl} to be an exact solution of string theory 
(at disk order) to all orders in $\alpha'$, it is not
clear why the fluctuations around it that have been considered 
in \cite{cama,lpt}
should reproduce the dynamics of the fundamental string far down the
spike. In the last section we will raise
some points about the relevance of this unsolved problem 
to the specific situation at hand.

With these cautionary remarks, we can see that it is not 
straightforward to
conclude that the results we find here apply directly to
fundamental strings. Though highly suggestive, 
it should be clear that we restrict ourselves to studying what
could be called BI-strings, i.e., those
described using exclusively the Born-Infeld action. 
We hope that future work will add further evidence that 
this description is also an appropriate one for fundamental strings.

\section{BI-string from the D2-brane}

The BI action describing a D2-brane and its coupling to a spacetime
3-form gauge potential $A$ is
\begin{eqnarray}\label{d2act}
I = -{1\over 4\pi^2 g}
\int d^3\xi &\bigg\{&\sqrt{-\det (\partial_\alpha X^\mu 
\partial_\beta X_\mu + 2\pi{\cal F}_{\alpha\beta})}\nonumber\\
&+&{1\over 6} \epsilon^{\alpha\beta\gamma}
A_{\mu\nu\rho} 
\partial_\alpha X^\mu \partial_\beta X^\nu \partial_\gamma X^\rho
\bigg\}.
\end{eqnarray}
Here $\mu,\nu,\rho=0,\dots,9$ are spacetime indices, and 
$\alpha,\beta,\gamma = 0,1,2$ are worldvolume indices. Also, $g$ 
is the (type IIA) string coupling constant and the units are
such that $\alpha'=1$.
We take the background spacetime to be flat, and the dilaton to be
constant.
The solutions of the BI theory we are interested in are most 
conveniently
analyzed by choosing the ansatz
\begin{eqnarray}\label{ansatz}
X^0 &=& t,\qquad\qquad 2\pi {\cal F}_{tz}={\cal E}(t,z),\nonumber\\
X^1 &=& z,\\
X^2 &=& R(t,z) \cos\sigma,\nonumber\\
X^3 &=& R(t,z) \sin\sigma.\nonumber
\end{eqnarray}
All other $X^i$ are taken
to be constant, while the remaining ${\cal F}_{\alpha\beta}$ 
are set to zero.
The coordinates $(t,z,\sigma)$ parametrize the worldvolume. We choose 
$0\leq \sigma <2\pi$, i.e., worldvolumes with cylindrical topology, 
but one should keep in mind that these circles of the D2-brane
are not wrapping any spacetime 1-cycle. The function 
$R(t,z)\leq 0$
measures the radius of these worldvolume circles, 
and worldvolume electric flux 
runs along $z$.
The background spacetime gauge field $H= dA$
will be taken to be uniform and ``aligned'' with the brane, 
$H_{0123}= h$.
The action reduces to
\begin{equation}\label{redact}
I = -{1\over 2\pi g}
\int dt\;dz \left( R \sqrt{1-{\dot R}^2 - {\cal E}^2 + {R'}^2}
- {h\over 2} R^2\right),
\end{equation}
where dots and primes denote derivatives with respect to $t$ and $z$,
respectively.

The worldvolume gauge field is more conveniently dealt with in terms
of the canonical momentum ${1\over 2\pi g}{\cal D}={\delta L/ 
\delta {\cal E}}$. 
Upon performing
a Legendre transformation the functional we use to describe the 
theory is
\begin{equation}\label{actd}
\tilde I[{\cal D},R] = -{1\over 2\pi g} \int dt\;dz \left(\sqrt{(R^2 + 
{\cal D}^2)
(1-{\dot R}^2+{R'}^2)}
- {h\over 2} R^2\right).
\end{equation}
The Gauss law constraint on the worldvolume gauge field translates into
$\partial_z {\cal D}= 0$. Moreover, the Euler-Lagrange equation for 
${\cal D}$ 
implies $\partial_t {\cal D}= 0$. 
Thus we conclude that ${\cal D}$ must be a constant. 

We will also need the energy function, which can be 
obtained straightforwardly by completing the passage
to Hamiltonian formalism. If we only want it for static solutions,
then we can read it
directly from (\ref{actd}) as
\begin{equation}\label{energ}
E = {1\over 2\pi g} \int dz \left(\sqrt{(R^2 + {\cal D}^2)(1+{R'}^2)}
- {h\over 2} R^2\right).
\end{equation}
Negative values of $h$ will oppose the expansion of the brane, 
so in the following we will restrict to $h\geq 0$.

The construction above can be easily generalized to any D$p$-brane,
by choosing $X^2,\dots,X^{p+1}$ to describe a $(p-1)$-sphere $S^{p-1}$
of radius $R(t,z)$. 
The D2-brane is somewhat particular in that there is an alternative
way to describe the system by starting from the eleven-dimensional 
membrane, or M2-brane.
In the latter, a non-vanishing string charge appears when 
the M2-brane
is wrapped around the compact eleventh dimension. In fact, this 
provides us with a simple way to determine the quantization condition
on the worldvolume electric field, as we show now.
The bosonic sector of the action is  
\begin{equation}\label{elm}
I_{(M2)} = -{1\over 4\pi^2}
\int d^3\xi \bigg\{\sqrt{-\det (\partial_\alpha {\bar X}^M 
\partial_\beta {\bar X}_M)}
+{1\over 6}\epsilon^{\alpha\beta\gamma}
A_{MNR} 
\partial_\alpha {\bar X}^M \partial_\beta {\bar X}^N 
\partial_\gamma {\bar X}^R\bigg\},
\end{equation}
where $M,N,R$ run over all eleven dimensional indices. 
Double-dimensional reduction of
this action leads to the Nambu-Goto action 
of the fundamental string, whereas direct reduction 
yields (\ref{d2act}) (see \cite{fourm} and references therein). 
In flat space, with a constant dilaton, the usual reduction of the 
eleven-dimensional metric to the string metric in ten dimensions 
requires us to scale ${\bar X}^\mu= g^{-1/3}X^\mu$. Besides, if the 
M2-brane wraps $n$ times the eleventh dimension we will have
\begin{equation}
{\bar X}^{11} = n\sigma r_{11} = n\sigma g^{2/3},
\end{equation}
where we have used the M-theory relation between the compact radius
$r_{11}$ and 
the string coupling $g$.
The reduced action that results from (\ref{elm}) is precisely equal 
to (\ref{actd}), with
\begin{equation}
{\cal D}= n g.
\end{equation}
Then we can say that there are $n$ fundamental strings dissolved in 
the D2-brane worldvolume.
Of course, the quantization condition can be derived
without recourse to the 
higher-dimensional interpretation with the same result, 
and can be extended to the other D$p$-branes.

\subsection{Static solutions}

Let us look for static solutions, i.e., those determined by extremizing
the energy (\ref{energ}) with respect to the field $R$,
\begin{equation}
{\delta E\over \delta R} - {d\over d z}
{\delta E\over \delta R'}= 0.
\end{equation}
This equation can be written as
\begin{equation}
{d\over dz}\left(\sqrt{R^2 + {\cal D}^2\over 1 + {R'}^2}- 
{h\over 2} R^2\right) =0.
\end{equation}
Thus we find, in terms of an integration constant $C$,
\begin{eqnarray}\label{statsol}
R' &=& -{\sqrt{ R^2 + {\cal D}^2 - 
(C+ h R^2/2)^2}\over C +h R^2/2}\nonumber\\
 &=& -{h\over h R^2 + 2C}\sqrt{(R^2_{max} - R^2)(R^2 - R^2_{min})},
\end{eqnarray}
where $R^2_{max,min}$ are simply defined in terms of $h$, ${\cal D}$,
$C$, by the second equation. From here we get $R(z)$
by integration. If $R'$ vanishes at some finite
$z=z_0$ (which happens if $R_{min}$ or 
$R_{max}$ are real and can be reached for finite $z$), 
then the solution admits
a natural continuation past $z_0$ by taking the other sign for 
the square root in (\ref{statsol}).

\begin{figure}
%\vspace{-0.5in}
\begin{center}\leavevmode  %
\epsfxsize=12cm\epsfbox{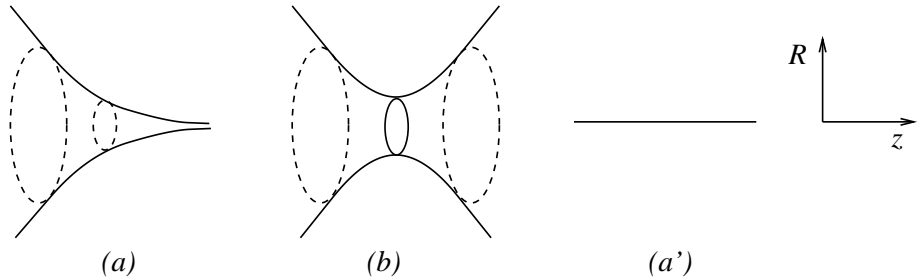}
\end{center}
%\vspace{-0.5in}
\caption{Static solutions of Born-Infeld theory (in the absence 
of spacetime background
field). A worldvolume electric field
runs along the $z$ direction on the D-brane.
(a) Spike. (b) Wormhole. (a') Born-Infeld string.
}
\label{fig:stats}
\end{figure}

In order to gain a little insight we shall first show how the solutions 
constructed in \cite{cama,bion}, in the absence of external fields, 
$h=0$, arise in our 
parametrization.\footnote{The ansatz and 
parametrization in \cite{cama,bion} is perhaps more
adequate to describe distortions from a flat D-brane, whereas we are 
interested in obtaining collapsed branes as well.} 
If $h=0$ then $R_{max} =\infty$ and the brane spans an infinite 
extension. 
Depending on the relative values of $C$ and ${\cal D}$, $R_{min}^2$ 
can be
zero, positive, or negative. The latter happens for $C<{\cal D}$, and
leads to solutions with
singularities at finite distances.\footnote{The Born-Infeld ``electron,''
or BIon \cite{bion}, corresponds to the limit $C=0$.} 
The other cases are (see fig.\ \ref{fig:stats})
 
\begin{enumerate}

\item[(a)] $C={\cal D}$, $R_{min}=0$: The ``spike'', or string 
ending on the brane. 
As $z\rightarrow\infty$, we have $R,R'\rightarrow 0$, and the 
brane becomes more
and more string-like. Explicitly, $R = R_0 e^{-z/{\cal D}}$, and 
(in contrast to higher $p$-branes) we never 
recover a flat D2-brane at finite $z$.

\item[(b)] $C>{\cal D}$, $R_{min}>0$: The wormhole solution. 
Now $R'$ vanishes at some finite $z=z_0$ when $R=R_{min}$. 
At this point, the solution must
be continued to the other sign for the root in (\ref{statsol}), 
yielding a wormhole throat symmetric about $z_0$.

\end{enumerate}

Case (a) has been argued in \cite{cama,bion,lpt} to be a BPS solution, 
and in \cite{thorl} to be an exact solution of string theory 
to all orders in $\alpha'$, at disk level. 
A degenerate solution directly related to this one is

\begin{enumerate}

\item[(a')] $C={\cal D}$, $R=R'=0$: The BI-string. This is a collapsed
cylinder and it can be viewed as the
limiting form of the spike as $z\rightarrow\infty$. The energy per
unit length of this 
configuration, i.e., the tension of the BI-string, is 
\begin{equation}
T_{BI}= {1\over 2\pi g} {\cal D}= n T_f,
\end{equation} 
where $T_f$ is the tension of a fundamental string. 

\end{enumerate}
Therefore, the BI-string has the right charge and tension to be 
identified with a number $n$ of fundamental strings. 
In what limit can this be reliable? Apart from $O(\alpha')$ 
corrections to the BI theory, which we do not address here and might be
relevant when departing from the BPS state,
we are neglecting the 
gravitational (closed string) effects of the string
while keeping its interaction with open strings. Hence 
we need $g^2 n\ll g n$, i.e., weak coupling $g\ll 1$.  
This is the same as requiring
the gravitational effects of the (single) D-brane to be small.
The effect is that we only keep disk
diagrams of string theory and neglect higher genera. 
In terms of the M2-brane, we need the compact radius $r_{11}$
to be small. 

Turn on now the external field, $h >0$. 
One consequence of having this background field is that
all supersymmetries are broken. Therefore, stability of the string
is no longer guaranteed, and in fact we will see later that it can 
decay.
As regards to the solutions (\ref{statsol}), the main change is 
that $R_{max}$ can now be finite and real. Besides, 
in order to be able
to neglect the self-gravitation of the field we have to require
$g h\ll 1$.

It is instructive to consider first the simplest solutions, 
with $R'=0$ for all $z$,
which describe cylindrical branes. Then there is just a single
degree of freedom, $R$, and the system is described by 
the potential energy,  
\begin{equation}\label{cylpot}
V(R) = {L\over 2\pi g} \left(\sqrt{R^2+ g^2 n^2} - 
{h\over 2} R^2\right).
\end{equation}
In order to find solutions with finite energies we have taken 
the direction $z$ to be compact with length $L$. The brane is 
therefore toroidal, but recall that one of the cycles of this 
torus, the one with radius $R$, does not wrap any non-trivial 
spacetime cycle. 

\begin{figure}
%\vspace{-0.5in}
\begin{center}\leavevmode  %
\epsfxsize=12cm\epsfbox{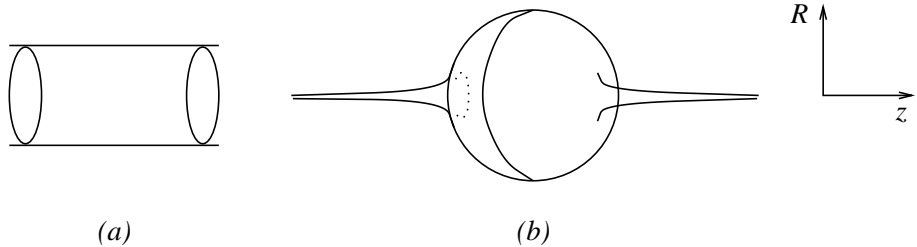}
\end{center}
%\vspace{-0.5in}
\caption{
Static, unstable configurations in the 
presence of a uniform
background 4-form field strength. The background field tends to pull
the D2-brane outwards and balances precisely the elastic tension.
}
\label{fig:sphal}
\end{figure}

The extrema of (\ref{cylpot}) describe classical solutions.
The BI-string corresponds to $R=0$, which is always an 
extremum. It is a local minimum for fields smaller than a 
critical value, $h<h_c 
= 1/(ng)$, so in these cases
the BI-string is classically stable. In particular, for $h=0$ the 
minimum is a global one, and the string is absolutely stable, as 
expected from
the fact that in the absence of the external field the BI-string is
supersymmetric. If $0<h<h_c$
the potential is minimized at $R=0$, then reaches
an (unstable) maximum at $R_s=h^{-1}\sqrt{1 - (n gh)^2}$, 
and afterwards, for larger $R$, $V(R)$ decreases without limit.
Such potential typically exhibits tunneling from the state at
$R=0$ (the BI-string) to a cylindrical brane with radius $R$
such that $V(R) = V(0)$. The 
cylinder of radius $R_s$ (identified to yield a torus), 
fig.\ \ref{fig:sphal}(a), 
is the sphaleron on top of the potential barrier.
This tunneling will be analyzed in the next subsection.
{}Finally, if $h\geq h_c$ the BI-string would 
become classically unstable. Notice that, for $n$ of order one, 
such fields are beyond the values for which gravitational 
effects can be neglected.

The solutions with non-trivial dependence on $z$ are more interesting. 
As before, static
solutions are possible only for background fields smaller than 
a certain critical value. If we choose $C={\cal D}$ then
the solutions tend to string-like configurations as 
$|z|\rightarrow\infty$, i.e., $R\rightarrow 
R_{min}=0$. In this case 
the integral in (\ref{statsol}) can be explicitly performed to find
\begin{equation}\label{boa}
\sqrt{R_{max}^2 - R^2} +{2n g\over h R_{max}}
\ln{R_{max} +\sqrt{R_{max}^2 - R^2}\over R} = |z - z_0|,
\end{equation}
with $R_{max} = {2\over h}\sqrt{1- n g h}$.

The first term on the left hand side of (\ref{boa}) would describe, 
if alone, a spherical brane. The second term, on the other hand, 
approaches a spike at large $|z|$. The complete solution 
describes a BI-string with a spheroidal bulge
centered at $z_0$, see fig.\ \ref{fig:sphal}(b). 
The solution ceases to exist for background fields larger 
than the same critical value $h_c$ we found above.

It is interesting to compute the energy of this configuration. Using 
(\ref{energ}), after some manipulation one finds the (exact) result
\begin{equation}\label{boaen}
E = \left({4\pi R_{max}^3 \over 3}\right) {h\over 8\pi^2 g} 
+ {n\over 2\pi}\int dz.
\end{equation}
The first contribution can be seen as a bulk energy
stored in a sphere of radius $R_{max}$, whereas the second
is the energy of $n$ fundamental strings. It is curious 
that such an exact 
split takes place, since both the string and the sphere are quite
distorted in the actual configuration. In contrast to the similar
split found for the spikes in \cite{cama}, supersymmetry can not be 
directly responsible for this, since it is completely broken.

\subsection{BI-string decay}

The static, unstable solutions we have found are properly 
interpreted as the 
sphalerons on top of a potential barrier, below which the BI-string
can tunnel quantum mechanically. The tunneling will be described by
instanton bounces in Euclidean time, so we set $t\rightarrow i\tau$.

It is easy to construct the instanton bounce solution describing
the tunneling to a cylindrical brane configuration, with $R'=0$. 
Again, in order to obtain
a finite decay rate the direction 
$z$ is taken to be compact with length $L$, and the brane will
be a right torus. 
Since the action is symmetric under exchange 
$\tau \leftrightarrow z$,\footnote{In fact, (\ref{redact}) is invariant
under boosts along $z$.} a
solution with $R'=0$ and $\partial_\tau R\neq 0$ can be found
by changing $z\rightarrow \tau$ in (\ref{boa}). 
Then it is clear that at $|\tau|\rightarrow \infty$
the solution reduces to the BI-string, while it bounces at the 
time-symmetric point $\tau=\tau_0$. This describes the tunneling
of a BI-string into a toroidal D2-brane of finite radius 
$R_{max}$. 
The subsequent
Lorentzian evolution consists of the expansion of the radius $R$, with 
almost uniform acceleration (for large radii).
The decay rate is, as usual, given to leading order by the exponential
of minus the classical
Euclidean action of the instanton, $\exp (-I_E)$. 
The computation of the
action is straightforward.  
We may want to use (the Euclidean counterpart of) the 
functional (\ref{actd}) instead of (\ref{redact}), 
since it implies that we keep fixed the 
number $n$ of strings. But then, in order to obtain the decay rate
we must subtract from the instanton action the action of the 
initial $n$ fundamental strings, call it $\tilde I_0$.
Taking this into account, in the end it does not make any 
difference whether we use the
functional $I$ in (\ref{redact}) 
(for which $I_0=0$), 
or $\tilde I -\tilde I_0$, using (\ref{actd}). 
It is evident that, for a toroidal solution, 
computation of (\ref{actd}) is
precisely equal, due to $\tau\leftrightarrow z$ symmetry,
to that of the energy of the static solution, 
(\ref{boaen}), from (\ref{energ}). 
Therefore, the decay rate of the string
into toroidal D2-branes is 
\begin{equation}
\Gamma_{tor} \sim \exp\left(-{Lh R_{max}^3 \over 6\pi g}\right). 
\end{equation}

It seems much more difficult to construct the instanton 
mediating the decay 
associated to the sphaleron (\ref{boa}). Nevertheless, it is
clear that the decay will proceed by nucleation of a spheroidal
bulge on the string, which will expand after tunneling. After 
subtracting from the action of the instanton
the contribution of the initial state, $\tilde I_0$, the decay rate
should be finite without any need of compactifying $z$.\footnote{It 
is nevertheless possible to find static solutions with spheroidal 
bulges periodic in $z$, which are also interpreted as sphalerons.} 
Even if we do not have the explicit solution, 
we know that if the external
field is small (precisely, if $h\ll (ng)^{-1}$), 
then the spheroidal bulge is very large and the
effect of the spikes is comparatively small. The decay rate is then
approximately equal to that of the nucleation of spherical 2-branes 
without worldvolume flux, computed in \cite{teit}.

The action for nucleating spheres and tori goes, respectively, 
like $\sim h^{-3}$ and $\sim L h^{-2}$, for $h$ much smaller
than the critical value. Thus, 
when the direction $z$ is infinite, or its periodic length $L$ is
very large compared to the 
length scale $h^{-1}$ imposed by the external field, then the decay
is dominated by nucleation of spheroidal bulges. When $L$ is of order
$h^{-1}$ or less, it becomes more favorable for the string to expand
into a toroidal brane.

{}Finally, notice that the presence of the string $(n\neq 0)$
involves a slight enhancement of the decay rate (and a decrease of
the energy of the sphalerons) 
as compared to the situation with branes in the absence of strings.

\section{Discussion}

As we mentioned above, it is easy to obtain a similar description of
D$p$-branes collapsed to strings for arbitrary $p\geq 2$
by replacing the $S^1$ in 
$X^2,X^3$ in (\ref{ansatz}) with a $S^{p-1}$ in $X^2,\dots,X^{p+1}$. 
In this way we can study the behavior of a 
string in a background RR $(p+2)$-form field strength.
The fact that for $p>3$ the D$p$-brane couples to 
a magnetic, instead of electric,
spacetime RR field, does not alter the analysis.
But it turns out that
there are qualitative differences between the results
for $p=2$ (the one analyzed above) and $p> 2$. 
In particular, the sphaleron solution
(\ref{boa}) does not exist for $p> 2$. In order to understand 
better the differences between $p=2$ and $p>2$ it is helpful to 
examine the potential $V_p(R)$ for cylindrical ($R'=0$) 
configurations of the brane. One finds
\begin{equation}\label{radpotp}
V_p(R) = {L \Omega_{p-1}\over (2\pi)^p g}
\left( \sqrt{R^{2p-2} + {\cal D}^2} - {h\over p} R^p\right),
\end{equation}
in the uniform background $H_{01\dots p+1}=h$, and where 
$\Omega_{p-1}$ is the volume of the unit $S^{p-1}$.
This potential is always extremized at $R=0$. But, for $p>2$, 
this is a local maximum for any $h>0$. Therefore 
the BI-string is rendered unstable even for arbitrarily weak 
fields. Indeed, for
small $h{\cal D}\sim hng$ and $p>2$ the potential 
$V_{p}(R)$ reaches a local minimum at
a finite (small) radius $\sim (hng)^{1\over p-2}$. For 
larger $R$, the shape of the potential is 
qualitatively similar to the one for $p=2$.
Nevertheless,
it is unclear whether this {\it classical} instability should 
be trusted: the
spacetime gauge and gravitational self-interactions of the brane
that we have been 
neglecting should be most important for small radii and would act so
that opposite sides of the cylinder would attract each other. 
This would tend to enhance
the classical stability of the brane.

As a matter of fact, string perturbation theory can be used, at least
in principle, to see if such instability is indeed present. 
This would imply studying disk diagrams with a closed string
insertion, with the boundary conditions on the disk chosen 
to describe D-branes with geometries such as considered in this paper, 
and the whole placed in the
spacetime background field $H$. The actual calculations, though,
present some technical complications, and this 
remains an interesting open problem.

Another extension of the work in this paper
involves considering magnetic fluxes, as opposed 
to electric, in the worldvolume of the D-brane. 
Then we would be describing D$p$-branes 
carrying the charge of D$(p-2n)$-branes \cite{mdoug}. For 
instance, a spherical 
D2-brane with magnetic flux distributed uniformly on its
worldvolume can carry the charge of a D0-brane \cite{town}, and, if
collapsed to a point, it would also have the
the right D0-brane mass. It would be very interesting to analyze
this specific problem within M(atrix) theory, where spherical membranes
have been recently studied \cite{ktr}.
By dualities, all these situations can be 
related to one of the above: with the ansatze we have been using,
the D$(p-2n)$-brane as a collapsed
D$p$-brane is described by essentially the same equations
as those for the string as a collapsed D$(2n+1)$-brane. 

Now we come to the question of stringy corrections to the Born-Infeld
action. Let us consider how they could affect
the effective radial potential (\ref{cylpot}). The latter contains
the relevant information about cylindrical solutions, and qualitative
guidance on the more generic situations where $R'\neq 0$. For large $R$
the stringy corrections should be small. But one could imagine
that at radii of order the string length, $\alpha'^{1/2}$, the potential 
might receive large corrections, e.g., perhaps it could develop a 
barrier that might prevent the occurrence of the tunneling process
we have described. However, such large corrections at certain values of
the radius $R$ appear as difficult
to reconcile with the fact that oscillatory fluctuations of the 
spike, which sweep all possible values of $R$, reproduce correctly the
behavior expected from string theory \cite{cama,lpt}. 
Although far from conclusive, this is suggestive that the short
distance corrections
to the BI description of the string may be kept under control. 
This may be particularly true for the specific case we have 
been considering in the paper, the D2-brane or 
equivalently the M2-brane. This is because the BI action of the 
wrapped M2-brane precisely reduces to the fundamental string action, 
even if for small $r_{11}$ the corrections to the BI description 
could have been relevant.

{}Finally, it would be interesting to be able to lift the 
restriction to weak coupling,
and include the spacetime gauge and gravitational backreaction
of the brane and the external field (the latter would become, far 
from the brane, a 
generalized Melvin-type field). This would imply generalizing 
the gravitational 
instantons for nucleation of branes that have been constructed 
in \cite{dggh}.

\section*{Acknowledgements}
We would like to thank Steve Giddings, 
Juan Maldacena and Amanda Peet for comments and useful conversations.
This work has been partially 
supported by FPI program (MEC-Spain) and by grant 
UPV 063.310-EB225/95.


\begin{thebibliography}{99}

\bibitem{joe} J. Polchinski, {\it TASI Lectures on D-branes,} 
hep-th/9611050.

\bibitem{leigh} R.G. Leigh, Mod. Phys. Lett. A4, 2767 (1989).

\bibitem{cama} C.G. Callan and J.M. Maldacena, {\it Brane Dynamics
From the Born-Infeld Action,} hep-th/9708147.

\bibitem{bion} G.W. Gibbons, {\it Born-Infeld Particles and 
Dirichlet p-branes,} hep-th/9709027.

\bibitem{lpt} S. Lee, A. Peet, and L. Thorlacius, {\it Brane-Waves
and Strings,} hep-th/9710097.

\bibitem{thorl} L. Thorlacius, {\it Born-Infeld String as a Boundary
Conformal Field Theory,} hep-th/9710181.

\bibitem{teit} C. Teitelboim, Phys. Lett. 167B, 63 (1986).

\bibitem{dggh} F. Dowker, J.P. Gauntlett, G.W. Gibbons and 
G.T. Horowitz,  Phys. Rev. D53, 7115 (1996), hep-th/9512154.

\bibitem{fourm} P.K. Townsend, {\it Four Lectures on M-Theory,}
hep-th/9612121.

\bibitem{mdoug} M.R. Douglas, {\it Branes Within Branes,}
hep-th/9512077.

\bibitem{town} P.K. Townsend, Phys. Lett. B373, 68 (1996), 
hep-th/9512062.

\bibitem{ktr} D. Kabat and W. Taylor, {\it Spherical Membranes in
Matrix Theory,} hep-th/9711078; S-J. Rey, {\it Gravitating M(atrix)
Q-Balls,} hep-th/9711081.

\end{thebibliography}
\end{document}